\documentclass[fleqn,usenatbib]{mnras}

\usepackage[T1]{fontenc}

\DeclareRobustCommand{\VAN}[3]{#2}
\let\VANthebibliography\thebibliography
\def\thebibliography{\DeclareRobustCommand{\VAN}[3]{##3}\VANthebibliography}

\usepackage{graphicx}	
\usepackage{amsmath}	
\usepackage{amssymb}	

\title[An evolutionary channel for CO-sdOs and He-sdOBVs]{An evolutionary channel for CO-rich and pulsating He-rich subdwarfs }

\author[Miller Bertolami et al.]{
M. M. Miller Bertolami,$^{1,2,3}$\thanks{E-mail: marcelo@MPA-Garching.MPG.DE, mmiller@fcaglp.unlp.edu.ar}
T. Battich,$^{3}$
A. H. C\'orsico,$^{1,2}$
L. G. Althaus,$^{1,2}$
F. C. Wachlin,$^{1,2}$
\\
$^{1}$Instituto de Astrof\'isica de La Plata, UNLP-CONICET, La Plata, Paseo del Bosque s/n, B1900FWA, Argentina\\
$^{2}$Facultad de Ciencias Astron\'omicas y Geof\'isicas, UNLP, La Plata, Paseo del Bosque s/n, B1900FWA, Argentina\\
$^{3}$Max-Planck-Institut f\"{u}r Astrophysics, Karl Schwarzschild Strasse 1, 85748, Garching, Germany
} 

\date{Accepted 30/02/2022. Received 31/11/2021; in original form 1BC}

\pubyear{2022}

\begin{document}

\label{firstpage}
\pagerange{\pageref{firstpage}--\pageref{lastpage}}
\maketitle

\begin{abstract}
Recently a new class of hot subluminous stars strongly enriched in C and O have been discovered (CO-sdOs). These stars show abundances very similar to those observed in PG1159 stars but at lower temperatures. Moreover, it has been recently suggested that C and O enrichment might be the key ingredient driving the pulsations in He-rich hot subdwarf stars (He-sdBVs). Here we argue that these two types of rare stars can be explained by a variant of one of the main channels forming hot subdwarf stars. The scenario involves the formation and merging of a He-core white dwarf and a less massive CO-core white dwarf.  We have constructed a simple merger models and computed their subsequent evolution. The merger products are in agreement with the surface parameters and composition of CO-sdOs. In addition, we have performed simulations including the effects of element diffusion and the excitation of pulsations. These simulations show that less massive merger products can form stellar structures that have surface parameters, abundances, and pulsation periods similar to those displayed by He-sdBVs. We conclude that the proposed scenario, or some variant of it, offers a very plausible explanation for the formation of CO-sdOs, pulsating He-sdBs and low-luminosity PG1159 stars.
\end{abstract}




\section{Introduction}

Hot subluminous stars of spectral types B and O (sdB, sdO) are late stages in the evolution of low-mass stars.
In the Hertzsprung–Russell diagram they are located between
the main sequence and the white-dwarf (WD) sequence, and correspond to the so-called Extreme Horizontal Branch (EHB) stars in globular clusters \citep{2009ARA&A..47..211H}. While most sdBs have atmospheres dominated by hydrogen (H), about 10\% of all hot subdwarf stars show
Helium(He)-dominated spectra and come in two spectral flavors; He-sdB and He-sdO \citep{2017A&A...600A..50G}. 
While the most common H-rich subdwarfs are supposed to be low-mass core
He-burning stars, the evolutionary status of the
He-rich subclass is less clear. He-rich subdwarfs have been suggested to be the result of either He~WD+He~WD mergers \citep{2000MNRAS.313..671S,2012MNRAS.419..452Z}, sdB+WD mergers \citep{2011MNRAS.410..984J}, or late helium core flashes in both single and binary evolution \citep{2001ApJ...562..368B,2008A&A...491..253M, 2018MNRAS.481.3810B}.

The population of H-rich sdBs harbors two families of pulsators; 
 the rapid pressure mode pulsators with periods ($P$) in the range $P \sim$~80--400~s discovered by \cite{1997MNRAS.285..640K} \cite[sdBVr, after][]{2010IBVS.5927....1K},  and the slow gravity(g) mode  
pulsators with periods $P \sim$~2500--7000~s discovered by \cite{2003ApJ...583L..31G} \cite[sdBVs, after ][]{2010IBVS.5927....1K}.
Pulsations in both groups of variable stars have been explained by the action of the
$\kappa$-$\gamma$ mechanism due to the partial ionization of iron group elements
in the outer layers, where these elements are enhanced by
the action of radiative levitation \citep{1997ApJ...483L.123C,2003ApJ...597..518F}. 

With the early discovery of g-mode pulsations in LS~IV-14°116 \citep{2005A&A...437L..51A}, with periods $P\sim 2200\mbox{--}3400$~s, and the later discoveries of similar pulsations in Feige~46   \citep{2019A&A...623L..12L}, $P\sim 1950\mbox{--}5100$~s, and PHL~417  \citep{2020MNRAS.499.3738O}, $P\sim 2280\mbox{--}6840$~s, it became clear that a small family of pulsators hides among the He-sdBs (from now on He-sdBVs). All these He-sdBVs have very similar surface compositions and photospheric parameters  ($X_{\rm H}\sim 0.36\mbox{--}0.48 $ and $X_{\rm He}\sim 0.4\mbox{--}0.64 $, by mass fractions, $\log g \sim 5.7\mbox{--}5.9$, and $\log T_{\rm eff}\sim 35\,500\mbox{--}36\,500$~K) that identifies them as part of the so called heavy-metal sdBs that show extreme zirconium- and lead-rich photospheres \citep[see][ and references therein]{2020MNRAS.491..874N}.

Explaining pulsations in He-sdBVs has been challenging. With temperatures around $T_{\rm eff}\sim 36\,000$~K, these stars are too hot for g-modes to be excited by the opacity bump created by the ionization of M-shell electrons in Ni and Fe.  \cite{2011ApJ...741L...3M}  proposed that pulsations in He-sdBVs could be explained by the $\epsilon$-mechanism \citep[see][]{1989nos..book.....U} during the He-core subflashes that appear in low-mass stellar models immediately before the star settles on the quiescent He-core burning stage (HeCB), i.e. during the pre-EHB stage. Although this mechanism predicts g-mode pulsators in the correct location of the $\log T_{\rm eff}$--$\log g$ diagram (from now on, Kiel diagram) it predicts periods which are short in comparison to those observed in He-sdBVs \citep[$P\sim $200--2000~s,][]{2018A&A...614A.136B}. This led these authors to explore the possibility that pulsations in He-sdBVs could be  stochastically excited by helium-flash-driven convection  during the same evolutionary stage \citep{2020NatAs...4...67M}. 
The main drawback of this scenario is that pulsations should be active only during the He-shell flashes themselves which corresponds to only 1\% of the pre-EHB stage. 
More recently \cite{2019MNRAS.482..758S} identified that pulsations in the right range of periods and at the right range of surface temperatures and gravities could be driven by the classical $\kappa$-$\gamma$ mechanism in stars with extreme amounts of carbon and oxygen enrichments; $X_{\rm C}\sim 0.2\mbox{--}0.3 $ and $X_{\rm O}\sim 0.2\mbox{--}0.3 $, by mass fractions. Such possibility would make He-sdBVs low-temperature counterparts of GW~Vir stars \citep[see][and references therein]{2019A&ARv..27....7C}. This is an extremely interesting possibility, albeit not devoided of shortcomings. First, He-sdBVs do not show significantly large amounts of C and O on their surfaces, which are completely dominated by H and He. As shown by \cite{2007ApJS..171..219Q}, in order for C and O to drive pulsations in that part of the 
Kiel diagram, the combined abundance of C and O needs to be higher than $\sim 0.4$ by mass fraction. Secondly, even if C and O could be kept hiding immediately below the photosphere it is necessary to identify an evolutionary scenario that can lead to an object with an extremely C-O enriched envelope in that part of the HR-diagram. The only known stellar objects with such CO-enhanced  envelopes are PG1159 stars \citep{2006PASP..118..183W} which are the consequence of late helium-shell flashes on the post-AGB phase (the so called "born again" scenario). This scenario predicts luminosities (surface gravities) more than 2 orders of magnitude higher (lower) for the effective temperatures typical of He-sdBVs, so it has to be discarded.  \cite{2019MNRAS.482..758S} suggested the possibility that binary interactions might remove $1.5 M_\odot$ from an already $2 M_\odot$  He main sequence, exposing 0.5$M_\odot$ CO-enhanced He-burning core. The absence of any binary signature in any of the three He-sdBV pulsators, makes the previous scenario doubtful. Even more, it demands for an explanation of how the He-main sequence star was formed and how binary interactions would manage to remove $1.5 M_\odot$ from a rather compact $2 M_\odot$ He-main sequence object without completely destroying it. Moreover, the presence of large quantities of H in the photospheres of these stars also suggests a different origin.

Recently, Werner et al. (this volume) discovered the existence of two CO-enhanced sdO stars (CO-sdOs): PG~1654+322 and PG~1528+025. These stars show abundances very similar to those observed in PG1159 stars with $(X_{\rm He}, X_{\rm C},X_{\rm O})\simeq (0.6, 0.15, 0.25)$ but at higher gravities, and located close to the He-main sequence. Due to their higher gravities these objects cannot be explained by the born again scenario. 

In this letter we present an evolutionary scenario that can explain the existence of both He-sdBVs and CO-sdOs, as well as explain the pulsations observed in the first group by means of the pulsation mechanism proposed by \cite{2019MNRAS.482..758S}. The proposed scenario is a natural consequence of one of the most accepted scenarios for the formation of hot subdwarf stars, and as such, fits nicely within our current understanding of their evolution and formation. 


\section{The evolutionary scenario}
\begin{figure}
\centering
	\includegraphics[width=6.5cm]{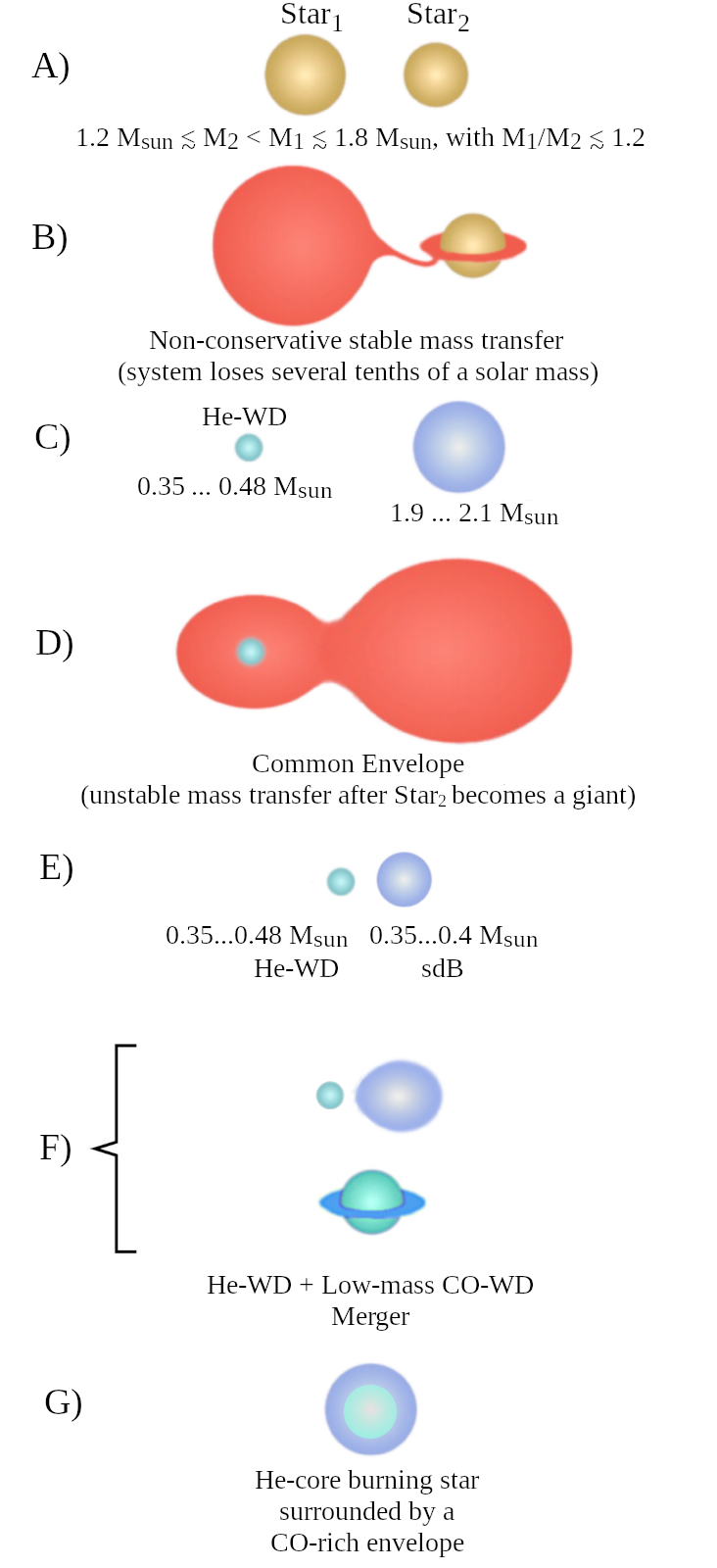}
    \caption{Evolutionary scenario for the formation of CO-sdOs and He-sdBVs. Initial masses are chosen so that the first mass transfer can be non-conservative ($M_1+M_2\gtrsim 2.6 M_\odot$) and stable ($M_1/M_2\lesssim$ 1.2), while allowing the formation of a binary with a massive He-WD ($\gtrsim 0.35 M_\odot$) and a $\sim 2M_\odot$ rejuvenated main sequence star.}
    \label{fig:scenario}
\end{figure}
\begin{figure}
	\includegraphics[width=\columnwidth]{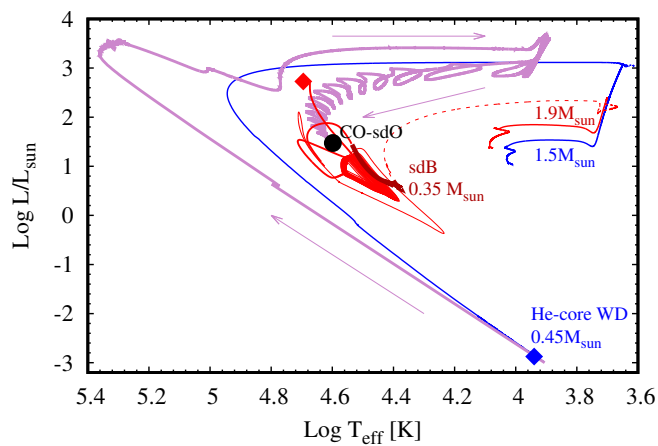}
    \caption{Hertzsprung-Russell diagram of the evolutionary sequences computed for our "toy merger" model. Blue tracks show the evolution of an initially 1.5$M_\odot$ star ($Z=0.001$) that fills its Roche lobe during the late RGB and ends as a $0.45 M_\odot$ He-core WD. Red tracks show the evolution of a  1.9 $M_\odot$ star (the rejuvenated Star$_2$ in stage c of Fig. \ref{fig:scenario}). The thick dark red line indicates the slow evolution during the He-core burning (sdB) stage. Thin red lines indicate the fast evolutionary stage during the thermal micropulses \citep{2007A&A...471..911G}. The red and blue diamonds indicate the moment at which the merger event happens. The purple line indicates the evolution after the merger event, arrows show the direction of the evolution. In the particular example shown here, the total CO-rich mass accreted by the $0.45 M_\odot$ He-core WD is $0.1 M_\odot$, leading to the formation of a $0.55 M_\odot$ CO-sdO (black circle). Dashed lines indicate the stages in which mass transfer is taking place.}
    \label{fig:toy}
\end{figure}

 \label{sec:scenario}
Population synthesis studies by \cite{2002MNRAS.336..449H,2003MNRAS.341..669H}
identified three main binary channels for forming sdB stars: (i) stable mass transfer (Roche lobe overflow, RLOF) followed by unstable mass transfer and common envelope evolution, (ii) an episode of stable RLOF, and (iii) the merger of two He-core WDs. The existence of short-period sdB binaries with WD companions, formed in the first channel (i), indicates that the first mass-transfer phase has to be both stable and very nonconservative \citep{2020A&A...635A.193G,2021ApJ...920...86K}. The evolutionary scenario proposed in this letter (see Fig. \ref{fig:scenario}) can be understood as a small subset\footnote{The fact that both He-sdBVs and CO-sdOs are very rare indicates that the channel leading to their formation has to correspond to a small part of the parameter space.} of channel (i). For most initial masses and semiaxes it is expected that channel (i) will form a pair in which the sdB is more massive than the WD companion or the WD companion is a CO-core WD. However, for a minority of systems one might expect the first mass transfer episode to begin close to the tip of the Red Giant Branch (TRGB), giving rise to rather massive He-core WD companions ($M_{\rm WD}\simeq 0.35\mbox{--}0.48 M_\odot$). Moreover, for a small group of these systems one should expect the initially less massive star (Star$_2$ in Fig. \ref{fig:scenario}) to increase its mass to $M_{\rm star 2}\simeq 1.9\mbox{--}2.1$ after the first mass transfer event. Stars in this mass range form the smallest He-burning cores and can form low-mass sdBs and CO-WDs \citep{2009A&A...507.1575P}, with masses $M_{\rm sdB}=M_{\rm CO-WD}\simeq 0.33\mbox{--}0.4 M_\odot$.
Two examples of these systems would correspond to a binary system (e.g. $Z=0.001$)  composed of a $M_1=1.5\,M_\odot$ primary and $M_2=1.3 \,M_\odot$ companion separated by $a\lesssim 0.796$ AU (alternatively $a\lesssim 0.289$ AU) and a period of $P\lesssim 155$d ($P\lesssim 34$d). After the end of the main sequence, the primary (Star$_1$) will evolve to the red giant branch and fills its Roche lobe at $R_l^1\lesssim 67 R_\odot$ ($R_l^1\lesssim 24 R_\odot$), when the He-core mass is $M\lesssim 0.45M_\odot$ ($M\lesssim 0.35M_\odot$). Mass transfer in that conditions is expected to be stable ($q=M_1/M_2 <1.2$) but mass transfer needs to be non-conservative. After this first mass transfer,  Star$_1$ will evolve, avoiding He ignition and forming a He-core WD. Assuming $M_2=1.9 M_\odot$ after the first mass transfer, for the system to undergo a common envelope event in stage D) the Roche lobe of Star$_2$ needs to be $R_l^2< 20 R_\odot$, which implies a separation in stage C) of $a\lesssim 0.18$AU and a period of $P\lesssim 19d$. This implies that the system needs to lose 70\% to 77\% of its initial orbital angular momentum. These numbers will vary with the choice of the metallicity of the models and their specific initial masses. Also, in realistic simulations it is possible that first mass transfer will start earlier than in these simple numerical examples lowering the amount of angular momentum to be lost. Still these examples highlight the need for the first mass transfer event to be very non-conservative as found by \cite{2002MNRAS.336..449H}. Despite the possible theoretical complications to explain the loss of angular momentum, we know from observations that these compact systems composed of a low-mass hot subdwarf with a low-mass WD companion exist \citep{2020A&A...635A.193G,2021ApJ...920...86K}.

Once the sdB ends its HeCB stage and evolves to the WD stage a merger might happen. The merger can happen either as a consequence of gravitational wave radiation when the star is already a cold CO-WD or just before the WD stage, when these low-mass stars undergo a series of flashes that increase their radii orders of magnitude, which will trigger a second unstable mass transfer event and a merger. The peculiarity of the proposed scenario is that the He-WD is the most massive and compact member of the pair. As a consequence, it is the sdB/CO-WD that will be disrupted and accreted on top of the He-WD, leading to the ignition of He in the core and the formation of a He-burning star with a CO-enriched envelope\footnote{ We expect the scenario to be also valid for He-WD with masses similar to that of the sdB ($M_1\sim M_2$). In these cases we expect the merger to lead to the destruction of both components and the complete mixing of their material and the formation of a He-core burning star with initially non-zero central CO abundances.}. Fig. \ref{fig:scenario} shows an schematic description of this scenario. These He-burning stars with CO-enriched envelopes should be located in the Kiel diagram close to the He-main sequence \citep{2012sse..book.....K}, which is exactly where both the He-sdBVs and the new CO-sdOs discovered by Werner et al. are located.  

\section{A toy merger model}
\label{sec:toy}
In order to test the scenario described in section \ref{sec:scenario} and Fig. \ref{fig:scenario} we have simulated with {\tt LPCODE} \citep{2005A&A...435..631A, 2016A&A...588A..25M} the evolution of the merger event using a simplified description of the process. We first constructed a 0.45 $M_\odot$ He-core WD by removing the envelope of an initially  1.5$M_\odot$ ($Z=0.001$) red giant branch star and followed its cooling on the white dwarf stage down to $\log L_{\rm WD}/L_\odot=-3$ (see Fig. \ref{fig:toy}). This corresponds to the evolution during stages A) to C) of Star$_1$ in Fig. \ref{fig:scenario}.  Following our previous example, we assume that Star$_2$ has an initial mass of $\sim 1.3 M_\odot$ and that $\sim 0.45 M_\odot$ are lost from the system during the non-conservative mass transfer. Then we evolved an initially $1.9 M_\odot$ star  (the rejuvenated Star$_2$ in stage C in Fig.  \ref{fig:scenario}) and removed most of its mass during the RGB, creating a low-mass sdB star (stage D of Fig.  \ref{fig:scenario}). This star is then evolved through the HeCB stage and, later, on the post-sdB and white dwarf stages (Fig. \ref{fig:toy}), this corresponds to stage E  of Fig.  \ref{fig:scenario}).

To model the merger process we proceed as follows. When the post-sdB object undergoes the He shell flashes immediate before the WD phase \cite[see][]{2009A&A...507.1575P} we assume that unstable mass tranfer starts, the post-sdB star is disrupted, forming a disk and then part of this material is accreted into the He-core WD (stages F and G in Fig.\ref{fig:scenario}). To estimate the composition of the accreted material we take the mean abundances (by mass) of the post-sdB model which are: $\langle X^{\rm post-sdB}_{\rm H}\rangle \simeq 4\times 10^{-3}$, $\langle X^{\rm post-sdB}_{\rm He}\rangle \simeq 0.237$, $\langle X^{\rm post-sdB}_{\rm C}\rangle \simeq 0.180$, $\langle X^{\rm post-sdB}_{\rm O}\rangle \simeq 0.578$.
With these abundances we then compute a rapid accretion process on the He-WD model ($\dot{M}=10^{-5} M_\odot$/yr). For this letter we have adopted four  different accreted masses $M_{\rm acc}=0.35, 0.25, 0.1,$ and $0.03 M_\odot$, ranging from full accretion of the CO-WD to 10\% accretion efficiency, this choice results in final merged objects of masses $M_{\rm merger}=0.8, 0.7, 0.55$ and $0.48 M_\odot$.

As soon as mass starts to be accreted into the He-WD, He is ignited in a shell, similarly to what happens in He-WD+He-WD mergers \citep{2000MNRAS.313..671S}. The accreting star then evolves back to the giant branch increasing its luminosity by 6 orders of magnitude (see the purple line in Fig. \ref{fig:toy}). After this initial expansion the merger product starts contracting and evolving towards the He-main sequence, as the model undergoes a series of He-shell subflashes that take place progressively deeper in the star, until steady He-burning begins (black circle in Fig. \ref{fig:toy}).
\begin{figure}
 \centering
	\includegraphics[width=7.5cm]{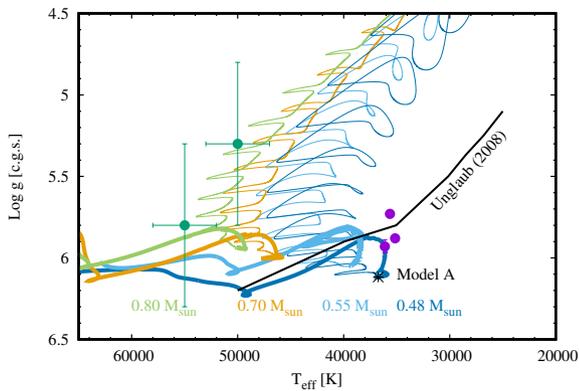}
    \caption{Kiel diagram showing the evolution of our four merger
products. Thin lines indicate the fast evolution before the
models settle on the HeCB, while thick lines show the slower
evolution during the HeCB and post-HeCB stage. Green and purple
circles indicate the location of the 2 known CO-sdOs and the 3
known He-sdBVs.  The black line indicate the wind limit derived by Unglaub (2008)
 for solar-like metallicities.  The
black star indicates the location of Model A from
Fig. \ref{fig:HesdBVs}.}
    \label{fig:kiel}
\end{figure}

\section{CO-rich hot subdwarf stars}

 \label{sec:evolution}
In the Kiel diagram the merger products settle during the steady HeCB stage close to the He main sequence and, consequently, their location is mostly determined by the final mass of the star (see Fig. \ref{fig:kiel}). The surface composition of these objects immediately after the merger event is set by the bulk composition of the disrupted star. In the case of our disrupted low-mass CO-WD these abundances are  
($\langle X_{\rm He}\rangle,\langle X_{\rm C}\rangle, \langle X_{\rm O}\rangle )^{\rm post-sdB} \simeq (0.237,  0.180, 0.578$), which are  similar to those recently observed by Werner et al. (this volume) in two CO-sdO stars  ($(X_{\rm He}, X_{\rm C},X_{\rm O})\simeq (0.6, 0.15, 0.25)$). Consequently He-, C- and O-rich surface compositions are a natural prediction of the current scenario. Significantly larger He abundances are expected in mergers with mass ratios close to 1, where both stars are expected to be disrupted and mixed. Also, it is expectable that mixing due to the inversion of the mean molecular weight will progressively decrease the CO content of the envelope. Interestingly, in this picture the mass of the star is directly correlated to its effective temperature. For the $\log T_{\rm eff}$ and $\log g$ values derived by Werner et al. a strong prediction of this scenario is that, within this picture the masses of the two CO-sdO discovered by Werner et al. (this volume) should be $M_{\rm CO-sdO}\simeq 0.7 \mbox{--}0.8 M_\odot$ (see Fig. \ref{fig:kiel}). 

Once stars enter the HeCB stage the photospheric composition of the stars will strongly depend on the competition between gravitational settling and winds \citep{2001A&A...374..570U, 2008A&A...486..923U}. 
 Unfortunately, there are no wind studies in this particular region of the HR diagram \citep{2002A&A...392..553V,2008A&A...486..923U} and for these peculiar surface compositions. In this context we can only make educated guesses of what the result of this competition between diffusion and winds will be. The black line in Fig. \ref{fig:kiel} shows the locus of the wind limit derived by \cite{2008A&A...486..923U}. For values of $\log g$ larger than this limit homogeneous winds might not be possible and H and He would be allowed to diffuse outwards, leading to the formation of H or He pure atmospheres. The wind limit derived by \cite{2008A&A...486..923U} was derived for a typical sdB mass of $0.5 M_\odot$ and typical solar-like surface compositions and, consequently, cannot be taken at face value in this discussion. However, one might argue that a star with a CO enriched surface as that arising from the merger scenario proposed here would be more able to sustain winds than those studied by \cite{2008A&A...486..923U}. If that is the case, the typical CO-enriched composition would remain visible throughout the whole HeCB stage explaining the recent CO-sdOs detected by Werner et al. (this volume) as well as some low-luminosity PG1159 stars that cannot be explained by the traditional born again scenario.

 \section{He-rich hot subdwarf pulsators}
 \label{sec:pulsations}
 \begin{figure}
 \centering
	\includegraphics[width=7.5cm]{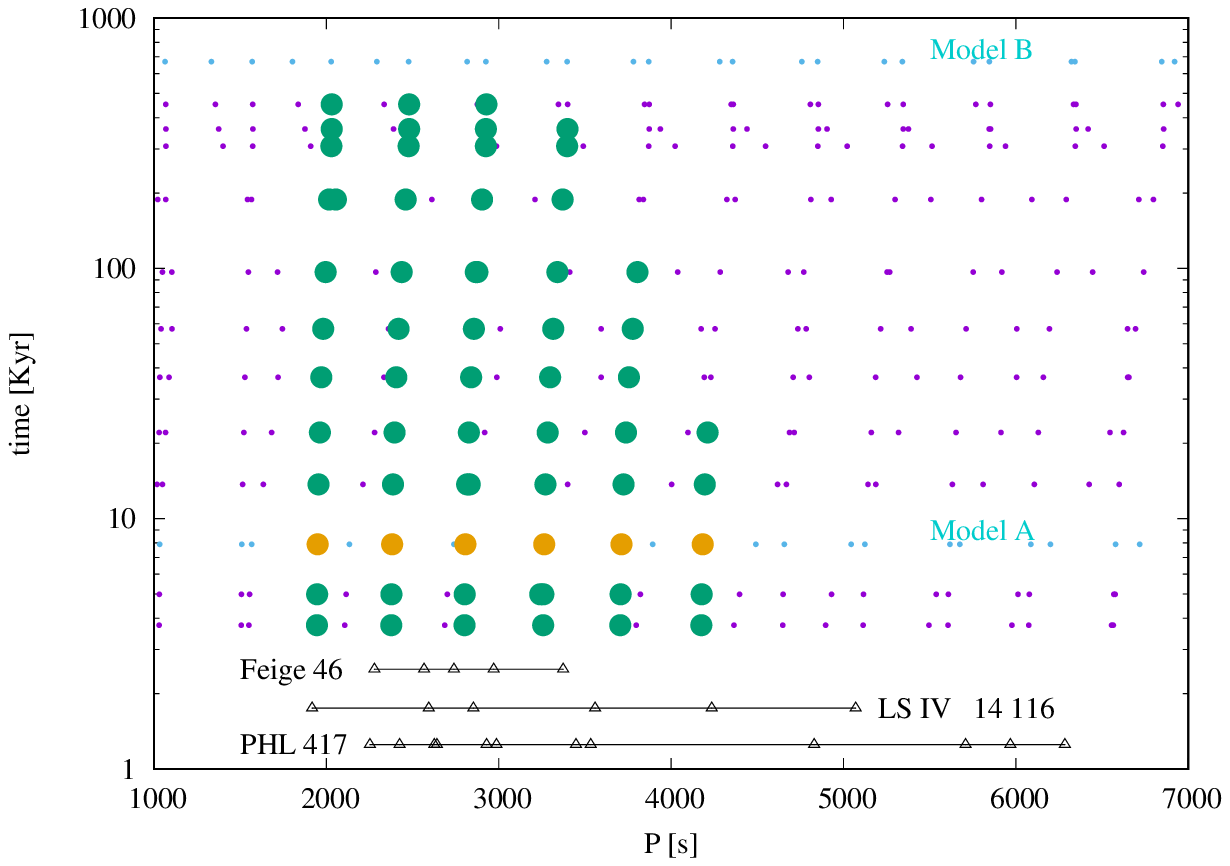}	
	\includegraphics[width=7.5cm]{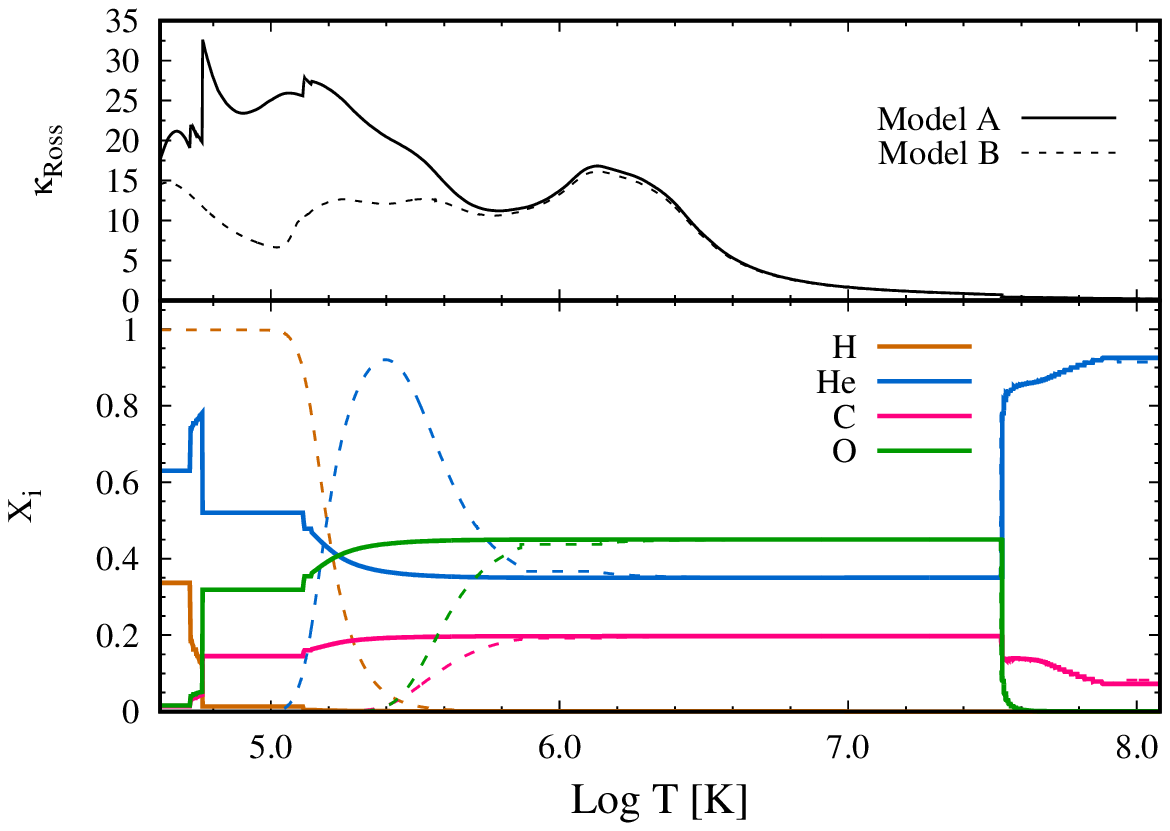}
    \caption{Upper panel: Filled circles indicate g modes that are excited by the $\kappa-\gamma$ mechanism in the models computed with diffusion ($M=0.48 M_\odot$). Orange circles indicate the modes excited in Model A. Dots indicate g modes that are not excited in the models. Open triangles indicate the observed pulsation modes in He-sdBVs. Middle panel: Rosseland mean opacity ($\kappa_{\rm Ross}$) of Models A and B discussed in the text. The driving of pulsations is provided by the opacity bump at $\log T\sim 6.2$. Lower panel: chemical profiles of models A and B discussed in the text.}
    \label{fig:HesdBVs}
\end{figure}
 
 Given the clear temperature dependence of the wind limit shown in Fig. \ref{fig:kiel}, one might expect winds to become less relevant as we go to less massive merger products, that settle at lower temperatures and luminosities. 
In fact,  merger models with $M_{\rm merger}<0.55 M_\odot$ settle on a region of the Kiel diagram where winds might not be possible (black line in Fig. \ref{fig:kiel}). While the wind limit derived by  \cite{2008A&A...486..923U} can not be taken at face value for CO-rich merger products, it is very interesting that pulsating He-rich sdBs are located very close to this limit. 
 
We have computed the evolution of our lower mass merger model ($0.48 M_\odot$) under the effect of diffusion and in the absence of winds. Then we have studied the excitation of g-modes with {\tt LP-PUL} non-adiabatic stellar pulsation code \citep{2006A&A...458..259C}. The results of these simulations can be seen in Fig. \ref{fig:HesdBVs}. As expected from the analysis of \cite{2019MNRAS.482..758S} we find that our models show excitation of g modes in the range of periods observed in the He-sdBV stars (upper panel of Fig. \ref{fig:HesdBVs}). As in the case of PG1159 stars these pulsations are mostly excited by the C and O opacity bumps at $\log T\sim 6.2$ (middle panel of Fig. \ref{fig:HesdBVs}). Even more interestingly, our models show that these pulsations are still excited when the surface abundances of the model have already been completely altered by diffusion. Model A in Fig. \ref{fig:HesdBVs} corresponds to a moment 8\,000 yr after the star settled on the HeCB. At that point the star has a H/He dominated surface composition of $X_{\rm H}=0.338$ and $X_{\rm He}=0.630$ with only traces of C and O ($X_{\rm C}=0.016$ and $X_{\rm O}=0.015$), see lower panel of  Fig. \ref{fig:HesdBVs}. Yet, as the composition of the inner part of the envelope is still unchanged by diffusion the $\kappa-\gamma$ mechanism is able to excite the full range of periods observed in Feige~46 and LS~IV~14$^\circ$116.  Consequently this model is able to explain, simultaneously, the location in the Kiel diagram, surface composition and pulsations observed in He-sdBVs. In our simulations, however, diffusion quickly removes the remaining He from the photosphere, turning the object into a normal sdB star (see discussion below). Diffusion is also responsible to stopping the pulsations in our models after $\sim 500\,000$ yrs (Fig. \ref{fig:HesdBVs}). After this time, the region at $\log T\sim 5.8$ becomes completely dominated by He. As a consequence of the removal of C and O, the damping of the pulsations in this region becomes more intense and pulsations cease. Interestingly, when this happens the driving of pulsations at 
 $\log T\sim 6.2$ is still active, yet unable to compensate for the additional damping at $\log T< 6.$ due to the increase in the He abundance. Model B in Fig. \ref{fig:HesdBVs} shows the chemical profile of the model at the moment of the cessation of the pulsations.

It is a well known fact that diffusion computations strongly overestimate the impact of diffusion in A- and B-type stars and that some other mechanisms have to reduce its efficiency. In the case, of He-sdBs candidates to slow down diffusion are turbulent mixing and winds. In fact, our models show the presence of several small convective zones near the photosphere of the star. If convective boundary mixing were able to connect these convective regions forming a deeper convective zone, this would slow down considerably the impact of diffusion at the photosphere. On the other hand, mixing due to the inversion of the mean molecular weight near the core might eventually decrease the CO content of the upper envelope and weaken the driving of pulsations.

\section{Conclusions}

We have presented a variant of one of the main evolutionary channels for the formation of sdB stars that is able to explain the new CO-rich sdO stars detected by Werner et al. (this volume), and low-luminosity PG1159. The scenario, sketched in Fig. \ref{fig:scenario}, involves the merging and disruption of a low-mass (pre) CO-WD, and the accretion of the material on top of a more massive He-WD companion. Although in the present letter we have only focused on the merging of a 0.45$M_\odot$ He-WD and a 0.35$M_\odot$ CO-WD (or pre-WD), similar results would arise from the merging of two low-mass CO-WDs  (0.35$M_\odot$)\footnote{Note that low mass CO-WDs have a significant helium mass content ($0.24$\% of the total mass) and should form a He-core burning star after the merger.}  or the merging of a He-WD and a  CO-WD of similar mass. In the two latter cases we expect the merger to lead to the direct collision and complete disruption of both components \citep{2014MNRAS.438...14D}, leading to somewhat different He/C/O compositions.

In addition to explaining the newly discovered CO-sdOs, we have shown that the current scenario (or some variant of it) is able to explain the properties observed in the He-rich sdB pulsators. Under the assumption that not all the mass of the companion is accreted on the He-WD, we have constructed models of different final masses. In particular models close to the lower mass end ($M_{\rm merger}\sim 0.5 M_\odot$) show pulsations driven by the classical $\kappa-\gamma$ mechanism that agree very well with the pulsations observed in He-sdBVs. Moreover, these models are located in a region of the HR-diagram where homogenous winds might not be possible. We find that, in the absence of winds, diffusion is able to quickly modify the surface composition of these stars, leading to surface compositions very similar to those of He-sdBVs without affecting the driving of pulsations for $\sim$500\,000 yr. After this time, however, He becomes dominant in the regions of the interior where $\log T\sim 5.8$ increasing the damping of the oscillations and finally killing them. The exact evolution of the surface abundances, as well as the length of time during which pulsations might be excited, will strongly depend on the interplay between mixing by Rayleigh–Taylor instabilities, winds, gravitational settling and the intensity of convective boundary mixing in the thin outer convective zones generated by the ionization of He, C and O. We plan to explore these topics in future works.

\section*{Acknowledgements}
The authors thank Klaus Werner and Nicole Reindl for sharing their preliminary results about the discovery of the CO-sdOs. This work was partially supported by PICT 2016-0053 from ANPCyT,  PIP 112-200801-00940 from CONICET, and grant G149 from University of La Plata. This research has made extensive use of NASA ADS.

\section*{Data Availability}
The data underlying this article will be shared on reasonable request to the corresponding author.



\bibliographystyle{mnras}
\bibliography{example} 









\bsp	
\label{lastpage}
\end{document}